\documentclass{llncs}
\newtheorem{cclaim}{Claim}
\usepackage{amssymb}
\title{Directed Feedback Vertex Set is Fixed-Parameter Tractable}
\author{Igor Razgon and Barry O'Sullivan\\
\{i.razgon,b.osullivan\}@cs.ucc.ie }
\institute{Computer Science
Department, University College Cork, Ireland}
\begin{document}
\maketitle

\begin{abstract}
We resolve positively a long standing open question regarding the
fixed-parameter tractability of the parameterized Directed Feedback
Vertex Set problem. In particular, we propose an algorithm which
solves this problem in $O(8^kk!*poly(n))$.
\end{abstract}
\section{Introduction}
In this paper we address the following problem. Given a directed
graph $G$ and a parameter $k$. Find a subset $S$ of vertices of $G$
of size at most $k$ such that any directed cycle of $G$ intersects
with $S$ or, if there is no such a subset, answer 'NO'. This is the
parameterized Directed Feedback Vertex Set (DFVS) problem. The
fixed-parameter tractability of this problem is a long-standing open
question in the area of parameterized complexity. In this paper we
resolve this question positively by proving the following theorem.

\begin{theorem} \label{main}
The parameterized DFVS problem can be solved in time
$O(8^kk!*poly(n))$ where $n$ is the number of vertices of $G$ and
$poly(n)$ is a polynomial on $n$ whose degree is a constant
independent of $k$.
\end{theorem}

\subsection{Overview of the Proposed Method}
First of all, we define a graph separation problem on a directed
\emph{acyclic} graph (DAG) $D$ as follows. Given two disjoint sets
$X=\{x_1, \dots, x_l\}$ and $Y=\{y_1, \dots, y_l\}$ of vertices  of
$D$ called the \emph{terminals}. A subset $R$ of non-terminal
vertices \emph{orderly separates} $X$ from $Y$ if $D \setminus R$
has no path from $x_i$ to $y_j$ for each $x_i,y_j$ such that $i \geq
j$. Find a subset $R$ as above of size at most $k$ or, if there is
no such a subset, answer 'NO'. We call this problem parameterized
\textsc{ordered multicut} in a DAG (\textsc{ord-mc-dag}). Now, the
proof of Theorem \ref{main} consists of two stages.

On the first stage we assume that the parameterized
\textsc{ord-mc-dag} problem is fixed-parameter tractable (FPT).
Under this assumption we prove that the parameterized DFVS problem
is FPT as well. In order to show this, we design an algorithm
solving the parameterized DFVS problem in time
$O(2^kk!*f(k,n)*poly(n))$, where $f(k,n)$ is the runtime of an
algorithm solving the parameterized \textsc{ord-mc-dag} problem. The
proposed algorithm for the parameterized DFVS problem is based on
the principle of iterative compression, which recently attracted a
considerable attention from researchers in the field
\cite{itercomp1,itercomp2,itercomp3}. The proposed algorithm appears
in \cite{RazgonDFVSORD} as a part of the proof that the
parameterized DFVS is FPT-equivalent to the parameterized
\textsc{ord-mc-dag} problem.

On the second stage we propose an algorithm solving the
parameterized \textsc{ord-mc-dag} problem in time $O(4^k*poly(n))$,
thus proving that the parameterized \textsc{ord-mc-dag} problem is
FPT. In order to design the algorithm we considered the
$O(4^k*poly(n))$ algorithm for the \textsc{multiway cut} problem on
\emph{undirected} graph proposed in \cite{ChenLiu}. The resulting
algorithm for the \textsc{ord-mc-dag} problem is obtained by
adaptation of the method proposed in \cite{ChenLiu} to the terms of
the \textsc{ord-mc-dag} problem.

Theorem \ref{main} immediately follows from combination of the above
two stages.

\subsection{Related Work}
Currently it is known that DFVS problem is FPT for a number of
classes of directed graphs \cite{ramantour,guotour,Gutin1}. These
classes are amenable to the \emph{short cycle} approach, according
to which a cycle of length $f(k)$ is identified and the branching is
performed on the vertices of the cycle with recursive invocation of
the algorithm to the corresponding \emph{residual} graph. However,
as noted in \cite{Gutin1}, the shortest cycle approach is unlikely
to lead to a parameterized algorithm for the general DFVS problem.

The connection between DFVS and the graph separation problem has
been noticed in \cite{DFVSapprox}, where a polynomial transformation
of DFVS to a version of the \textsc{multicut} problem on directed
graphs has been described. This connection has been refined in
\cite{RazgonDFVSORD} where the parameterized \textsc{ord-mc-dag}
problem has been introduced and proven to be FPT- equivalent to the
parameterized DFVS problem. As said in the previous subsection, a
part of the proof serves as the first stage of the proof of Theorem
\ref{main} of the present paper.

There has been a considerable attention from the parameterized
complexity community to the separation problems on \emph{undirected}
graphs. FPT-algorithms for the \textsc{multiway cut} problem and a
restricted version of the \textsc{multicut} problem were proposed in
\cite{Marxcut}. An improved algorithm for the \textsc{multiway cut}
problem has been proposed in \cite{ChenLiu}. As mentioned above, an
adaptation of this algorithm to the \textsc{ord-mc-dag} problem
serves as the second stage of the proof of Theorem \ref{main}.
Improved algorithms solving the \textsc{multicut} problem for a
number of special classes of graphs are proposed in
\cite{Niedercut}.

For the parameterized DFVS problem on \emph{undirected} graphs, the
challenging questions were to design an algorithm solving this
problem in $O(c^k*poly(n))$ where $c$ is a constant and to obtain a
polynomially bounded \emph{kernel} for this problem. The former
problem has been solved independently in \cite{itercomp1,NiederFVS},
the size of the constant has been further improved in
\cite{ChenFomin}. The latter problem has been solved first in
\cite{FellowsFVS}. The size of the kernel has been drastically
improved in \cite{BodlaenderFVS}.

Finally, non-trivial exact exponential algorithms for non-directed
and directed FVS problems appear in
\cite{Razgon1,Fomin4,RazgonICTCS}.

\subsection{Notations}
Let $G$ be a directed graph. We denote its sets of vertices and
edges by $V(G)$ and $E(G)$, respectively. Let $(u,v) \in E(G)$. Then
$(u,v)$ is a \emph{leaving} edge of $u$ and an \emph{entering} edge
of $v$. Accordingly, $u$ is an \emph{entering neighbor} of $v$ and
$v$ is a \emph{leaving} neighbor of $u$. Also, $u$ is the
\emph{tail} of $(u,v)$ and $v$ is the \emph{head} of $(u,v)$. A
vertex $u$ is \emph{minimal} if it has no entering neighbors and
\emph{maximal} if it has no leaving neighbors.

Let $ES \subseteq E(G)$. We denote by $G[ES]$ the subgraph of $G$
created by the edges of $ES$ and the vertices incident to them. We
denote by $G \setminus ES$ the graph obtained from $G$ by removal of
the edges of $ES$. For a set $R \subseteq V(G)$, $G \setminus R$
denotes the graph obtained from $G$ by removal the vertices of $R$
and their incident edges

In our discussion we frequently mention a \emph{path}, a
\emph{cycle}, or a \emph{walk} in a directed graph. By default, we
mean that they are \emph{directed} ones.

A directed feedback vertex set (DFVS) of $G$ is a subset $S$ of
$V(G)$ such that $G \setminus S$ is a directed acyclic graph (DAG).
Let $A$ and $B$ be disjoint subsets of vertices of $V(G)$. A set $R
\subseteq V(G) \setminus (A \cup B)$ \emph{separates} $A$ from $B$
if $G \setminus R$ has no path from any vertex of $A$ to any vertex
of $B$.

The parameterized problems considered in this paper get as input an
additional parameter $k$ and their task is to find an output of size
at most $k$ or to answer 'NO' if there is no such an output. A
parameterized problem is fixed-parameter tractable (FPT) if it can
be solved in time $O(g(k)*poly(n))$, where $n$ is the size of the
problem (in this paper, the number of vertices of the underlying
graph), $poly(n)$ is a polynomial on $n$ whose degree is a constant
independent of $k$. Sometimes we call the time $O(g(k)*poly(n))$ an
\emph{FPT-time} and an algorithm solving the given problem in an
FPT-time an \emph{FPT-algorithm}.

\subsection{Organization of the paper}
The rest of the paper is a proof of Theorem \ref{main}. Section 2
presents the first stage of the proof and Section 3 presents the
second stage of the proof as outlined in the above overview.

\section{Parameterized DFVS problem is FPT if Parameterized ORD-MC-DAG problem is FPT}
Let $D$ be a DAG and let $X=\{x_1, \dots, x_l\}$, $Y=\{y_1, \dots,
y_l\}$ be two disjoint subsets of its vertices called \emph{the
terminals}. We say that a subset $R$ of non-terminal vertices of $D$
\emph{orderly separates} $X$ from $Y$ if $D \setminus R$ has no path
from $x_i$ to $y_j$ for all $i$, $j$ from $1$ to $l$ such that $i
\geq j$. We call the corresponding problem of finding the smallest
set of non-terminal vertices orderly separating $X$ from $Y$
\textsc{ordered multicut} in a DAG and abbreviate it as
\textsc{ord-mac-dg}. \footnote{For the sake of convenience of the
analysis, we admit some abuse of notation treating sets as ordered
sequences. To circumvent this problem we can consider that the
vertices are assigned with names so that $(x_1, \dots, x_l)$ is the
lexicographic ordering of the names of $X$ and $(y_1, \dots, y_l)$
is the lexicographic ordering of the names of $Y$.} The
\emph{parameterized} \textsc{ord-mc-dag} problem gets as an
additional parameter an integer $k \geq 0$, its task is to find a
set $R$ orderly separating $X$ from $Y$ of size at most $k$ or to
say 'NO' if there is no such a set.  In this section we assume that
the parameterized \textsc{ord-mc-dag} problem is FPT and let
$SolveORDMCDAG(D,X,Y,k)$ be a procedure solving this problem in an
FPT-time. Based on this assumption, we design an FPT-algorithm for
the parameterized DFVS problem.

The proposed algorithm for DFVS is based on the principle of
\emph{iterative compression} which recently proved successful for
the design of parameterized algorithms for a number of problems. In
particular, let $v_1, \dots, v_n$ be the vertices of the input graph
$G$. The algorithm iteratively generates a sequence of graphs $G_0,
\dots, G_n$ where $G_0$ is the empty graph and $G_i$ is the subgraph
of $G$ induced by $\{v_1, \dots, v_i\}$. For each generated graph
the algorithm maintains a DFVS $S_i$ of this graph having size at
most $k$ or returns 'NO' if for some $G_i$ it turns out to be
impossible. If the algorithm succeeds to construct $S_n$ it is
returned because this is a DFVS of $G=G_n$ having size at most $k$.

The sets $S_i$ are computed recursively. In particular, $S_0=
\emptyset$. For each $S_i$, $i>0$, if $S_{i-1}$ is a DFVS for $G_i$
then $S_i=S_{i-1}$. Otherwise, if $|S_{i-1}| \leq k-1$, then
$S_{i}=S_{i-1} \cup \{v_i\}$. Finally, if none of the above two
cases is satisfied then we denote $S_{i-1} \cup \{v_i\}$ by $S'_i$
(observe that $|S'_i|=k+1$) and try to get a DFVS $S_i$ of $G_i$ of
size smaller than $S'_i$. In particular, for each subset $F$ of
$S'_i$, the algorithm applies procedure $ReplaceDFVS(G_i \setminus
F, S'_i \setminus F)$ whose output is a DFVS $F'$ of $G_i \setminus
F$ of size smaller than $S'_i \setminus F$ and disjoint with $S'_i
\setminus F$ or 'NO' if none exists. If we succeed to find at least
one such $F'$ then $S_i=F \cup F'$. Otherwise, 'NO' is returned. In
other words, the algorithm guesses all possibilities of $F=S'_i \cap
S_i$ and for each guessed set $F$ the algorithm tries to find an
appropriate set $S_i \setminus S'_i$. Clearly the desired set $S_i$
exists if and only if at least one of these attempts is successful.

The pseudocode of the $ReplaceDFVS$ function is shown below.
\\
\\
$ReplaceDFVS(G,S)$\\
Parameters: a directed graph $G$ and a DFVS $S$ of $G$, $|S|$ denoted by $m$.\\
Output: a DFVS $R$ of $G$ which is disjoint with $S$ and having size
smaller than $S$ or 'NO' if no such $R$ exists.
\begin{enumerate}
\item {\bf If} $G$ is acyclic {\bf then} return the empty set. %1
\item {\bf If} $S$ induces cycles {\bf then} return 'NO'. %2
\item Let $ES$ be the set of all edges of $G$ entering to the
vertices of $S$. %3
\item {\bf For each} possible ordering $s_1, \dots, s_m$ of the vertices of $S$ {\bf do} %4
\item~~~~~For each $s_i$, let $T_i$ be the set of vertices $w$ of $G \setminus S$ such that
$G[ES]$ has a path from $w$ to $s_i$. %7
\item~~~~Let $G'$ be a graph obtained from $G \setminus ES$ by introducing a set
$T=\{t_1, \dots, t_m\}$ of new vertices and for each $t_i$
introducing an edge $(w,t_i)$ for each $w \in T_i$ \footnote{Note
that $G \setminus ES$ is a DAG because any cycle of $G$ includes a
vertex of $S$ and hence an edge of $ES$. By construction, $G'$ is
DAG as well. Note also that graphs $G'$ are isomorphic for all
possible orders, we introduce the operation within the cycle for
convenience only.}
\item~~~~{\bf If} $SolveORDMCDAG(G',S,T,|S|-1)$
does not return 'NO' {\bf then} return the output of
$SolveORDMCDAG(G',S,T,|S|-1)$
\item {\bf endfor}
\item Return 'NO'
\end{enumerate}

Denote by $f(k,n)$ the time complexity of $SolveORDMCDAG$ applied to
a graph of $n$ vertices and parameter $k$ and let us evaluate the
time complexity of the above algorithm for the parameterized DFVS
problem. For each of $n$ iterations, the algorithm checks at most
$2^{k+1}$ subsets of vertices of the current DFVS. Each check
involves the run of the $ReplaceDFVS$ function with the size of its
second parameter bounded by $k+1$. Accordingly, the number of
distinct orderings explored by the main cycle of the function is at
most $(k+1)!$ For each ordering, the function $SolveORDMCDAG$ is
called exactly once and the size of its last parameter is bounded by
$k$. The resulting runtime is $O(2^k*k!*f(k,n)*poly(n))$, where
$poly(n)$  takes into account the $O(n)$ iterations of the iterative
compression method, auxiliary operations such as checking whether
the given set is indeed a DFVS of $G$, and factor $k+1$ of the above
factorial.

The non-trivial part of the analysis is the correctness proof of
$ReplaceDFVS$, which is provided by the following theorem.

\begin{theorem}
If $ReplaceDFVS(G,S)$ returns a set $R$, it satisfies the output
specification and conversely, if 'NO' is returned, then there is no
set satisfying the output specification.
\end{theorem}

{\bf Proof.} Assume first that $ReplaceDFVS(G,S)$ returns a set $R$.
This means that there is an ordering $s_1, \dots, s_m$ of $S$ such
that $R$ orderly separates $S$ from $T$ in $G'$ where $T$ and $G'$
are as defined by the algorithm. By definition of an orderly
separating set, $R \subseteq V(G) \setminus S$. Assume by
contradiction that $R$ is not a DFVS of $G$ and let $C$ be a cycle
of $G \setminus R$.

By definition of $ES$, the graph $G \setminus ES$ is acyclic
therefore $C$ contains edges of $ES$. Partition the edges of $ES$ in
$C$ into maximal paths. Let $P_1, \dots, P_l$ be these paths listed
by the order of their appearance in $C$. It follows from definition
of $ES$ that each $P_i$ ends with a vertex $s_{j_i}$ for some $j_i$.
Since line 2 of $ReplaceDFVS(G,S)$ rules out the possibility that
the edges of $ES$ may induce cycles and due to the maximality of
$P_i$, path $P_i$ begins with a vertex  which does not belong to $S$
that is, with some $w_i \in T_{j_i}$. Considering again that $G[ES]$
is acyclic, in order to connect $P_1, \dots, P_l$ into a cycle, $C$
includes a path in $G \setminus R \setminus ES$ from $s_{j_1}$ to a
vertex of $T_{j_2}$, $\dots$ , from $s_{j_{l-1}}$ to a vertex of
$T_{j_l}$, from $s_{j_l}$ to $T_{j_1}$. Clearly $(j_1 \geq j_2) \vee
\dots \vee (j_{l-1} \geq j_l) \vee (j_l \geq j_1)$ because otherwise
we get a contradictory inequality $j_1 < j_1$. Thus $G \setminus R
\setminus ES=(G \setminus ES) \setminus R$ has a path from some
$s_i$ to a vertex of $T_j$ such that $i \geq j$. By definition of
$G'$, graph $G' \setminus R$ has a path from $s_i$ to $t_j$ in
contradiction to our assumption that $R$ orderly separates $S$ from
$T$ in $G'$. This contradiction proves that $R$ is a DFVS of $G$.

Now, consider the opposite direction. We prove that if $R$ is a DFVS
of $G$ disjoint from $S$ and of size at most $|S|-1$ then it orderly
separates $S$ from $T$ in $G'$ for \emph{at least one ordering}
$s_1, \dots, s_m$ of $S$. It will immediately follow that if
$SolveORDMCDAG$ function returns 'NO' for \emph{all} possible orders
then there is no DFVS of $G$ with the desired property and the
answer 'NO' returned by $ReplaceDFVS(G,S)$ in this case is valid.

So, let $R$ be a DFVS of $G$ with the desired properties and fix an
arbitrary ordering $s_1, \dots, s_m$ of $S$. Let $t_1, \dots, t_m$
and $G'$ be as in the description of $ReplaceDFVS(G,R)$. Then the
following two claims hold.

\begin{cclaim} \label{claim1}
For each $i$, $G' \setminus R$ has no path from $s_i$ to $t_i$.
\end{cclaim}

{\bf Proof.} Assume that this is not true and let $P$ be such a
path, let $w$ be the immediate predecessor of $t_i$ in this path. By
definition of $G'$, the prefix $P''$ of $P$ ending by $w$ is a path
of $G \setminus R$. Taking into account the definition of $G'$, $w
\in T_i$ and $G$ has a path $P'$ from $w$ to $s_i$ including the
edges of $ES$ only. Observe that the vertices of $P'$ do not
intersect with $R$. Really, the heads of all edges of $P'$ belong to
$S$ which is disjoint from $R$ by definition, the first vertex $w$
does not belong to $R$ because $w$ participates in a path of $G
\setminus R$. Thus path $P'$ is a subgraph of $G \setminus R$. The
concatenation of $P'$ and $P''$ creates a closed walk in $G
\setminus R$, which, of course, contains a cycle obtained by taking
the closest repeated vertices. This is a contradiction to our
assumption that $R$ is a DFVS of $G$. $\square$

\begin{cclaim} \label{claim2}
Fix an arbitrary $l$ such that $1 \leq l \leq m$. Then there is $p$
such that $1 \leq p \leq l$ such that $G' \setminus R$ no path from
$s_p$ to any other $t_i$ from $1$ to $l$.
\end{cclaim}

{\bf Proof.} Intuitively, the argument we use in this proof is
analogous to the argument one uses to demonstrate existence of
minimal vertices in a DAG.

Assume that the claim is not true. Fix an arbitrary $i$, $1 \leq i
\leq l$. Since according to claim 1, $G' \setminus R$ has no path
from $s_i$ to $t_i$, there is some $z(i)$, $1 \leq z(i) \leq l$,
$z(i) \neq i$ such that $G' \setminus R$ has a path $P_i$ from $s_i$
to $t_{z(i)}$.

Consider a sequence $i_0, \dots, i_l$, where $i_0=i$,
$i_j=z(i_{j-1})$ for each $j$ from $1$ to $l$. This is a sequence of
length $l+1$ whose elements are numbers from $1$ to $l$. Clearly
there are at least two equal elements in this sequence. We may
assume w. l. o. g. that these are elements $i_0$ and $i_y$ where $1
\leq y \leq l$ (if these elements are $i_q$ and $i_r$ where $0 <q
<r$ we can just set $i_0=i_q$ and rebuild the above sequence). For
each $j$ from $0$ to $y-1$, consider the path $P'_{i_j}$ obtained
from path $P_{i_j}$ by removal of its last vertex. By definition of
$G'$, $P'_{i_j}$ is a path in $G \setminus R$ and finishing by a
vertex $w_{i_{j+1}} \in T_{i_{j+1}}$.

Let $P''_1, \dots, P''_y$ be paths in $G[ES]$ such that each $P''_j$
is a path from $w_{i_j}$ to $s_{i_j}$ (such a path exists by the
definition of $w_{i_j}$). Arguing as in Claim \ref{claim1}, one can
see that each $P''_j$ is a path in $G \setminus R$. Consequently, $G
\setminus R$ has a directed walk obtained by the following
concatenation of paths: $P'_{i_0},P''_1, \dots,P'_{i_{y-1}},P''_y$.
This walk begins with $s_{i_0}$ and finishes with $s_{i_y}$. Since
we assumed that $i_0=i_y$, we have a closed walk in $G \setminus R$
which contains a cycle in contradiction to the definition of $R$ as
a DFVS of $G$. $\square$

Now, we construct the desired ordering by a process that resembles
the topological sorting. Fix an index $p$ such that $s_p$ does not
have a path to any $t_i$ in $G'$ as guaranteed by Claim
\ref{claim2}. If $p \neq m$ then interchange $s_p$ and $s_m$ in the
ordering being constructed (of course if two terminals of $S$
interchange, then the corresponding terminals of $T$, $t_p$ and
$t_m$ in the considered case, interchange as well). Assume that the
last $m-l$ vertices in the ordering of $S$ have been fixed. If $l=1$
then, taking into account that $G' \setminus R$ has no path from
$s_1$ to $t_1$ in $G' \setminus R$ by Claim \ref{claim1}, the
resulting ordering is ready. Otherwise, fix $p$, $1 \leq p \leq l$
as stated by Claim \ref{claim2}. If $p \neq l$, interchange $s_l$
and $s_p$ in the ordering. Proceed until all the elements of the
order are fixed. $\blacksquare$

Thus, in this section we have proved the following theorem.

\begin{theorem} \label{part1}
The parameterized DFVS problem can be solved in time of
$O(2^k*k!*f(k,n)*poly(n))$, where $f(k,n)$ is the time of solving
the parameterized \textsc{ord-mc-dag} problem on a graph with $O(n)$
vertices.
\end{theorem}

\section{Parameterized ORD-MC-DAG problem is FPT}
In this section we provide an FPT algorithm for the parameterized
\textsc{ord-mc-dag} problem whose input is a DAG $G$, the sets
$X=\{x_1, \dots, x_l\}$ and $Y=\{y_1, \dots, y_l\}$ of terminals,
and a parameter $k \geq 0$. First of all, we notice that we may
assume that all vertices of $X$ are minimal ones and all vertices of
$Y$ are maximal ones. In particular, we show that graph $G$ can be
efficiently transformed into a graph $G'$, $V(G)=V(G')$, for which
this assumption is satisfied so that a set $R$ orderly separates $X$
from $Y$ in $G$ if and only if $R$ orderly separates $X$ from $Y$ in
$G'$.

Let $G'$ be a graph obtained from $G$ by the following 2-stages
transformation. On the first stage, remove all entering edges of
each $x_i$ and all leaving edges of each $y_i$. On the second stage
we introduce new edge $(u,v)$ for each pair of non-terminal vertices
$u,v$ such that $G$ has edges $(u,x_i),(x_i,v)$ or $(u,y_i),(y_i,v)$
for some terminal $x_i$ or $y_i$ (of course, new edges are
introduced only for those pairs that do not have edges $(u,v)$ in
$G$). Let $G'$ be the resulting graph. Note that $G'$ is a DAG
because it is a subgraph of the transitive closure of $G$.

\begin{proposition} \label{assumption}
A set $R \subseteq V(G) \setminus (X \cup Y)$ orderly separates $X$
from $Y$ in $G$ if and only if it orderly separates $X$ from $Y$ in
$G'$.
\end{proposition}

{\bf Proof.} Assume that $R$ orderly separates $X$ from $Y$ in $G$
but does not do this in $G'$ and let $P$ be a path from $x_i$ to
$y_j$ ($i \geq j$) in $G' \setminus R$. Replace each edge $(u,v)$
which is not present in $G$ by the pair of edges of $G$ which are
replaced by $(u,v)$ according to the above transformation. The
resulting sequence $P'$ of vertices form a walk in $G$. Since $G$ is
a DAG, vertex repetitions (and cycles as a result) cannot occur,
hence $P'$ is a path in $G$. The vertices of $V(P') \setminus V(P)$
are terminal ones, hence they do not belong to $R$. Consequently,
$P'$ is a path from $x_i$ to $y_j$ in $G \setminus R$, in
contradiction to our assumption regarding $R$.

Assume now that $R$ has the orderly separation property regarding
$G'$ but fails to orderly separate the specified pairs of terminals
in $G$. Let $P$ be a path from $x_i$ to $y_j$ in $G \setminus R$
such that $i \geq j$. Replace each appearance of an intermediate
terminal vertex in $P$ by an edge from its predecessor to its
successor in $P$. As a result we obtained a path from $x_i$ to $y_j$
in $G' \setminus R$ in contradiction to our assumption.
$\blacksquare$

Proposition \ref{assumption} justifies the validity of our
assumption that the vertices of $X$ are minimal in $G$ and the
vertices of $Y$ are maximal ones.

In order to proceed, we extend our notation. We denote by
$OrdSep(G,X,Y)$ the size of the smallest set of vertices of $G
\setminus (X \cup Y)$ orderly separating $X$ from $Y$ in $G$. If
$(x_i,y_j) \in E(G)$ for some $i$ and $j$ such that $i \geq j$, we
set $OrdSep(G,X,Y)=\infty$ because even the removal of all
nonterminal vertices will not orderly separate $X$ from $Y$. For two
disjoint subsets $A$ and $B$ of $V(G)$, we denote by $Sep(G,A,B)$
the size of the smallest subset of $V(G) \setminus (A \cup B)$
separating $A$ from $B$. If for some $u \in A$ and $v \in B$, $(u,v)
\in E(G)$ we set $Sep(G,A,B)=\infty$. If $A$ consists of a single
vertex $u$, we write $Sep(G,u,B)$ instead $Sep(G,\{u\},B)$. We
denote by $G^C(u)$ the graph obtained from $G$ by removal of $u$ and
adding all possible edges $(u_1,u_2)$ such that $u_1$ is an entering
neighbor of $u$, $u_2$ is a leaving neighbor of $u$ and there is no
edge $(u_1,u_2)$ in $G$.

The method of solving the \textsc{ord-mc-dag} problem presented
below is an adaptation to the \textsc{ord-mc-dag} problem of the
algorithm for the \textsc{multiway cut} problem in undirected graphs
\cite{ChenLiu}. In particular, the following theorem, which is the
cornerstone of the proposed method, is an adaptation of Theorem 3.2.
of \cite{ChenLiu}.

\begin{theorem} \label{leaveneighb}
Assume that $OrdSep(G,X,Y)< \infty$. Let $u$ be a leaving neighbor
of $x_l$ and assume that $Sep(G,x_l,Y)=Sep(G^C(u),x_l,Y)$. Then
$OrdSep(G,X,Y)=OrdSep(G^C(u),X,Y)$.
\end{theorem}

{\bf Proof.} Let $S_m$ be the set of vertices of $G^C(u) \setminus
(X \cup Y)$ of size $Sep(G^C(u),x_l,Y)$ which separates $x_l$ from
$Y$ in $G^C(u)$. Observe that $S_m$ separates $x_l$ from $Y$ in $G$.
Really, let $P$ be a path from $x_l$ to some $y_j$ in $G$. If it
does not include $u$ then the same path is present in $G^C(u)$,
hence it includes a vertex of $S_m$. Otherwise, $P$ includes $u$.
Since $OrdSep(G,X,Y)< \infty$, $u \notin Y$, hence it has a
predecessor $u_1$ and a successor $u_2$. It follows that $G^C(u)$
has a path obtained from $P$ by removing $u$ and adding edge
$(u_1,u_2)$, this new path includes a vertex of $S_m$, hence $P$
itself does.

Consider the graph $G \setminus S_m$. Let $C_1 \subseteq V(G
\setminus S_m)$ including $x_l$ and all the vertices reachable from
$x_l$ in $G \setminus S_m$. Let $C_2$ be the rest of vertices of $G
\setminus S_m$. Note that $u \in C_1$ because otherwise $u \in S_m$
in contradiction to our assumption.

Let $S_k$ be the smallest subset of vertices of $V(G) \setminus (X
\cup Y)$ that orderly separates $X$ from $Y$ in $G$. The sets
$C_1,S_m,C_2$ impose a partition of $S_k$ into sets $A=S_k \cap
C_1$, $B=S_k \cap S_m$ and $C=S_k \cap C_2$.

Consider now the graph $G \setminus C_1$. Let $S'_m$ be the subset
of $S_m$ consisting of vertices $v$ such that $G \setminus C_1$ has
a path from $v$ to some $y_j$ which does not include any vertex of
$B \cup C$. We are going to prove that $|S'_m| \leq |A|$.

Since $S_m$ separates $x_l$ from $Y$ in $G$ and is a smallest one
subject to this property (by the assumption of the lemma), $G$ has
$|S_m|$ internally vertex-disjoint paths from $x_l$ to $Y$ each
includes exactly one vertex of $S_m$ (by Menger's Theorem). Consider
the prefixes of these paths which end up at the vertices of $S_m$.
As a result we have a subset ${\bf P}$ of $|S_m|$ internally
vertex-disjoint paths, each starts at $x_l$ ends up at a distinct
vertex of $S_m$. Consider the subset ${\bf P'}$ of those $|S'_m|$
paths of ${\bf P}$ which end up at the vertices of $S'_m$.

Observe that each of these paths includes a vertex of $A$. Really
let $P_1$ be a path of ${\bf P'}$ which does not include a vertex of
$A$. Let $s$ be the final vertex of $P_1$. Observe that all vertices
of $P_1$ except $s$ belong to $C_1$: as witnessed by $P_1 \setminus
s$ they are reachable from $x_l$ by a path that does not meet any
vertex of $S_m$. Since $B$ and $C$ are subsets of $C_2$, $P_1
\setminus s$ does not intersect with $B$ and $C$. Let $P_2$ be a
path in $G \setminus C_1$ from $s$ to $y_j$ which does not include
the vertices of $B$ and $C$, which exists by definition of $S'_m$.
Taking into account that $A \subseteq C_1$, $P_2$ does not include
the vertices of $A$ as well. Let $P$ be the concatenation of $P_1$
and $P_2$. Clearly, $P$ is a path (vertex repetition is impossible
in a DAG) from $x_l$ to $y_j$ which intersects with neither of $A$,
$B$, $C$, that is, it does not intersect with $S_k$ in contradiction
to the fact that $S_k$ orderly separates $X$ from $Y$ in $G$. Thus
we obtain that $|S'_m| \leq |A|$.

Consider now the set $S'_k=S'_m \cup B \cup C$. By definition,
$|S'_k|=|S'_m|+|B|+|C|$ and $|S_k|=|A|+|B|+|C|$. Taking into account
that $|S'_m| \leq |A|$ as proven above, it follows that $|S'_k| \leq
|S_k|$. As well, $u \notin S'_k$ just because $S'_k$ does not
intersect with $C_1$. We are going to prove that $S'_k$ orderly
separates $X$ from $Y$ in $G$, which will finish the proof of the
theorem.

Assume by contradiction that this is not so and consider a path $P$
from $x_i$ to $y_j$ in $G \setminus S'_k$ such that $i \geq j$.
Assume first that $P$ does not intersect with $C_1$. That is, $P$ is
a path of $G \setminus C_1$. Since $S_k$ orderly separates $X$ and
$Y$, $P$ includes at least one vertex of $S_k$ or, more precisely,
at least one vertex of $V(G \setminus C_1) \cap S_k=B \cup C$. This
means that $P$ includes at least one vertex of $S'_k$ in
contradiction to our assumption.

Assume now that $P$ includes a vertex $w$ of $C_1$. By definition,
there is a path $P_1$ from $x_l$ to $w$ in $G \setminus S_m$. Let
$P_2$ be the suffix of $P$ starting at $w$. The concatenation of
$P_1$ and $P_2$ results in a path $P'$ from $x_l$ to $y_j$. By
definition, this path must include vertices of $S_m$ and, since
$P_1$ does not intersect with $S_m$, $P_2$ does. Let $s$ be the
\emph{last} vertex of $S_m$ which we meet if we traverse $P_2$ from
$w$ to $y_j$ and consider the suffix $P''$ of $P_2$ starting at $s$.

Observe that $P''$ does not intersect with $C_1$ because this
contradicts our assumption that $s$ is the last vertex of $P_2$
which belongs to $S_m$. Really, if there is a vertex $v \in C_1 \cap
P''$, draw a path $P_3$ from $x_l$ to $v$ which does not include any
of $S_m$, take the suffix $P_4$ of $P''$ starting at $v$,
concatenate $P_3$ and $P_4$ and get a path from $x_l$ to $y_j$ which
implies that $P_4$ must intersect with $S_m$ (because $P_3$ cannot)
and a vertex $s'$ of this intersection is a vertex of $P''$. Since
$s \notin C_1$, $v \neq s$, that is $v$ is a successor of $s$ in
$P''$, so is $s'$. Since $s \neq s'$ (to avoid cycles), $s'$ is a
vertex of $S_m$ occurring in $P''$, and hence in $P_2$, later than
$s$, in contradiction to the definition of $s$.

Thus $P''$ belongs to $G \setminus C_1$. Since $P''$ is a suffix of
$P$ which does not intersect with $S'_k$, $P''$ does not intersect
with $S'_k$ as well, in particular, it does not intersect with $B
\cup C$. It follows that $s \in S'_m$ in contradiction to the
definition of $P$. $\blacksquare$

Below we present an FPT-algorithm for the \textsc{ord-mc-dag}
problem. The algorithm is presented as a function
$FindCut(G,X,Y,k)$.
\\
\\
$FindCut(G,X,Y,k)$
\begin{enumerate}
\item {\bf If} $|X|=1$ {\bf then} compute the output efficiently. %
\item {\bf If} $Sep(G,x_l,Y)>k$ {\bf then} return 'NO' %
\item {\bf If} $x_l$ has no leaving neighbors {\bf then} return
$FindCut(G \setminus \{x_l,y_l\},X \setminus \{x_l\},Y \setminus
\{y_l\},k)$ (i.e., orderly separate $x_1, \dots, x_{l-1}$ from $y_1, %4
\dots, y_{l-1}$)
\item Select a leaving neighbor $u$ of $x_l$ %5
\item {\bf If}
$Sep(G^C(u),x_l,Y)=Sep(G,x_l,Y)$ {\bf then} return
$FindCut(G^C(u),X,Y)$.
\item Let $S_1=FindCut(G \setminus u, X,Y,k-1)$ and
$S_2=FindCut(G^C(u),X,Y,k)$. If $S_1 \neq 'NO'$, return $\{u\} \cup
S_1$. Else, if $S_2 \neq 'NO'$, return $S_2$. Else, return 'NO'.
\end{enumerate}

Before we provide a formal analysis of the algorithm, note the
properties of the \textsc{ord-mc-dag} problem that make it amenable
to the proposed approach. The first useful property is that vertex
$x_l$ has to be separated from \emph{all} the vertices of $Y$. This
property ensures the correctness of Theorem \ref{leaveneighb} and
makes possible ``shrinking" of the problem if the condition of Step
5 is satisfied. The second property is that if the condition of step
3 is satisfied, i.e. the vertices $x_l$ and $y_l$ are of no use
anymore, then, as a result of their deletion, we again obtain an
instance of the \textsc{ord-mc-dag} problem, i.e. we can again
identify a vertex of $X \setminus \{x_l\}$ to be separated from all
the vertices of $Y \setminus \{y_l\}$ and hence Theorem
\ref{leaveneighb} applies again.

In order to analyze the algorithm we introduce a definition of a
\emph{legal input}. A tuple $(G,X,Y,k)$ is a legal input if $G$ is a
DAG, $X$ and $Y$ are subsets of $V(G)$, the vertices of $X$ are
minimal, the vertices of $X$ are maximal, $|X|=|Y|$, $k \geq 0$.
Since $FindCut$ is initially applied to a legal input, the following
lemma proves correctness of $FindCut$.

\begin{lemma} \label{correct2}
Let $(G,X,Y,k)$ be a legal input with $|X|=l$. Then
$FindCut(G,X,Y,k)$ returns a correct output in a finite amount of
recursive applications. Moreover, all tuples to which $FindCut$ is
applied recursively during its execution are legal inputs.
\end{lemma}

{\bf Proof.} The proof is by induction on $|V(G)|$. In the smallest
possible legal input, graph $G$ consists of 2 vertices $x_1$ and
$y_1$, $X=\{x_1\}$, $Y=\{y_1\}$. According to the description of the
algorithm, this is a trivial case which is computed correctly
without recursive application of $FindCut$. The rest of the proof is
an easy, though lengthy, verification of the lemma for all cases of
recursive application of $FindCut$.

Assume now that $|V(G)|>2$. If $l=1$ or $Sep(G,x_l,Y)>k$, the output
is correct according to the description of the algorithm (the
correctness of the latter case follows from the obvious inequality
$Sep(G,x_l,Y) \leq OrdSep(G,X,Y)$). If $x_l$ has no leaving
neighbors then $FindCut$ is recursively applied to the tuple $(G
\setminus \{x_l,y_l\},X \setminus \{x_l\},Y \setminus \{y_l\},k)$.
Clearly, this tuple is a legal input, hence the lemma holds
regarding this input by the induction assumption, in particular the
output of $FindCut(G \setminus \{x_l,y_l\},X \setminus \{x_l\},Y
\setminus \{y_l\},k)$ is correct. Since $x_l$ has no leaving
neighbors, it has no path to the vertices of $Y$. Hence, any subset
of vertices orderly separating $X \setminus \{x_l\}$ from $Y
\setminus \{y_l\}$, orderly separates $X$ from $Y$ and vice versa.
It follows that the output of $FindCut(G \setminus \{x_l,y_l\},X
\setminus \{x_l\},Y \setminus \{y_l\},k)$ is a correct output of
$FindCut(G,X,Y,k)$ and hence the lemma holds regarding $(G,X,Y,k)$.

Assume that the algorithm selects such a leaving neighbor $u$ of
$x_l$ such that $Sep(G,x_l,Y)=Sep(G^C(u),x_l,Y)$. Then $FindCut$ is
recursively applied to $(G^C(u),X,Y,k)$. Observe that $u$ is a
non-terminal vertex because if $u=y_i$ ($u$ cannot be $x_i$ because
all the vertices of $X$ are minimal ones) then
$Sep(G,x_l,Y)=\infty>k$ and 'NO' would be returned on an earlier
stage. It follows that $(G^C(u),X,Y,k)$ is a legal input.  Taking
into account that $|V(G^C(u))|<|V(G)|$, the lemma holds regarding
$(G,X,Y,k)$ by the induction assumption, in particular, the output
$R$ of $FindCut(G^C(u),X,Y,k)$ is correct. Assume that $R \neq
'NO'$. Then $R$ is subset of non-terminal vertices of size at most
$k$, which orderly separates $X$ from $Y$ in $G^C(u)$. Assume that
$R$ does not orderly separate $X$ from $Y$ in $G$. Then $G \setminus
R$ has a path $P$ from $x_i$ to $y_j$ such that $i \geq j$. If $P$
does not include $u$ then this path is present in $G^C(u)$.
Otherwise, taking into account that $u$ is non-terminal vertex, this
path can be transformed into a path in $G^C(u)$ by removal $u$ and
introducing edge $(u_1,u_2)$ where $u_1$ and $u_2$ are the immediate
predecessor and the immediate successor of $u$ in $P$, respectively.
In both cases $P$ intersects with $R$, a contradiction. This
contradiction shows that $R$ orderly separates $X$ from $Y$ in $G$.
If $FindCutG^C(u),X,Y,k)$ returns 'NO' this means that
$OrdSep(G^C(u),X,Y)>k$. By Theorem \ref{leaveneighb}, in the
considered case $OrdSep(G^C(u),X,Y)=OrdSep(G,X,Y)$, that is
$OrdSep(G,X,Y)>k$ and hence the answer 'NO' returned by
$FindCut(G,X,Y)$ is correct. It follows that the lemma holds for the
considered case.

Assume now that none of the previous cases holds. In this case the
algorithm selects a leaving neighbor $u$ of $x_l$ such that
$Sep(G,x_l,Y)<Sep(G^C(u),x_l,Y)$ and applies itself recursively to
$(G \setminus u,X,Y,k-1)$ and $(G^C(u),X,Y,k)$. Observe that $u$ is
not a terminal vertex because if $u=y_i$ ($u$ cannot be $x_i$
because all the vertices of $X$ are minimal ones) then
$Sep(G,x_l,Y)=\infty>k$, hence an earlier condition is satisfied.
Note also that $k>0$. Really if $k=0$ then $Sep(G,x_l,Y)=0$ to avoid
satisfaction of an earlier condition. But this means that there is
no path from $x_l$ to the vertices of $Y$ hence either $x_l$ has no
leaving neighbors or for any leaving neighbor of $u$,
$Sep(G^C(u),x_l,Y)=Sep(x_l,Y)=0$, in any case one of the earlier
conditions is satisfied. It follows that both $(G \setminus
u,X,Y,k-1)$ and $(G^C(u),X,Y,k)$ are legal inputs. Since the graphs
involved in these inputs have less vertices than $G$, the recursive
applications of $FindCut$ to these tuples are correct by the
induction assumption. Assume that the output $R$ of $FindCut(G
\setminus u,X,Y,k-1)$ is not 'NO'. Then $R$ is a set of nonterminal
vertices of size at most $k-1$ which separates $X$ from $Y$ in $G
\setminus u$. Clearly that $R \cup \{u\}$ returned by
$FindCut(G,X,Y,k)$ in this case is correct. Assume now that
$FindCut(G \setminus u,X,Y,k-1)$ returns 'NO'. Clearly this means
that there is no subset $R$ separating $X$ and $Y$ in $G$ such that
$|R| \leq k$ and $u \in R$. Assume in this case that the output $R$
of $FindCut(G^C(u),X,Y,k)$ is not 'NO'. Arguing as in the previous
paragraph, we see that $R$ orderly separates $X$ from $Y$ in $G$,
hence the output $R$ returned by $FindCut(G,X,Y,k)$ in the
considered case is correct. Finally assume that
$FindCut(G^C(u),X,Y,k)$ returns 'NO'. Clearly, this means that there
is no subset $R$ of non-terminal vertices orderly separating $X$
from $Y$ in $G$ such that $|R| \leq k$ and $u \notin R$. Thus, any
decision regarding $u$ does not result in getting the desired
orderly separating subset. Hence, such a subset does not exist and
the answer 'NO' returned by $FindCut(G,X,Y,k)$ in the considered
case is correct. $\blacksquare$

Lemma \ref{correct2} allows us to define a search tree whose nodes
are associated with the legal inputs to which $FindCut(G,X,Y,k)$ is
recursively applied during its execution. The root of the tree is
associated with $(G,X,Y,k)$. Let $(G',X',Y',k')$ be a node of this
tree where $X'=\{x'_1, \dots, x'_{l'}\}$, $Y'=\{y'_1, \dots,
y'_{l'}\}$ (for convenience we identify a node with the tuple
associated with this node). If $FindCut(G',X',Y',k')$ does not apply
itself recursively then $(G',X',Y',k')$ is a leaf. Otherwise,
depending on the particular branching decision, $(G',X',Y',k')$ has
the child $(G' \setminus \{x'_{l'},y'_{l'}\}, X' \setminus
\{x'_{l'}\},Y' \setminus \{y'_{l'}\})$ or the child
$(G'^C(u),X',Y',k')$ or children $(G' \setminus u, X',Y',k'-1)$ and
$(G'^C(u),X',Y',k')$, where $u$ is a leaving neighbor of $x'_{l'}$.

\begin{lemma} \label{numleaves}
The number $L(G,X,Y,k)$ of leaves of the tree rooted by $(G,X,Y,k)$
is $O(4^k)$.
\end{lemma}

{\bf Proof.} For the legal input $(G,X,Y,k)$ with $|X|=l$, let
$m=max(2k+1-Sep(G,x_l,Y),0)$. We are going to prove that the number
of leaves of the search tree is at most $2^m$. Taking into account
that $m \leq 2k+1$, the result will immediately follow.

The proof is by induction on the number $N(G,X,Y,k)$ of nodes of the
tree rooted by $(G,X,Y,k)$. If $N(G,X,Y,k)=1$ then, taking into
account that $m \geq 0$, the statement immediately follows. Consider
the situation where $N(G,X,Y,k)>1$. Assume first that $(G,X,Y,k)$
has exactly one child $(G',X',Y',k)$ with $|X'|=l'$. Clearly
$L(G,X,Y,k)=L(G',X',Y',k)$. Let $m'=max(2k+1-Sep(G',x_{l'},Y'),0)$.
Observe that $m' \leq m$. Really, if $(G',X',Y',k)=(G^C(u),X,Y,k)$,
then $m'=m$ by the description of the algorithm. Otherwise,
$(G',X',Y',k)=(G \setminus \{x_l,y_l\},X \setminus \{x_l\},Y
\setminus \{y_l\},k)$. This type of child is created only if
$Sep(G,x_l,Y)=0$. Clearly, in this case $m' \leq m$. Taking into
account the induction assumption, we get $N(G,X,Y,k)=N(G',X',Y',k)
\leq 2^{m'} \leq 2^m$, as required.

Consider the case where $(G,X,Y,k)$ has two children $(G \setminus
u,X,Y,k-1)$ and $(G^C(u),X,Y,k)$ where $u$ is a leaving neighbor of
$x_l$. Observe that in this case $m>0$. Really, if $m=0$ then
$Sep(G,x_l,Y)>k$ which corresponds to an earlier non-recursive case.
Thus $m=2k+1-Sep(G,x_l,Y)$. Let $m_1=max(2(k-1)+1-Sep(G \setminus
u,x_l,Y),0)$. Taking into account that $Sep(G \setminus u,x_l,Y)
\geq Sep(G,x_l,Y)-1$, $m_1 < m$. Let
$m_2=max(2k+1-Sep(G^C(u),x_l,Y),0)$. By the description of the
algorithm, $Sep(G^C(u),x_l,Y)>Sep(G,x_l,Y)$, hence $m_2<m$. We
obtain $L(G,X,Y,k)=L(G \setminus u,X,Y,k-1)+L(G^C(u),X,Y,k) \leq
2^{m_1}+2^{m_2} \leq 2^{m-1}+2^{m-1}=2^m$, the second inequality
follows by the induction assumption. $\blacksquare$

According to Lemma \ref{correct2}, each node $(G',X',Y',k')$ of the
search tree is a valid input and hence $|V(G')| \geq 2$. On the
other hand if $(G',X',Y',k')$ is a non-leaf node and
$(G'',X'',Y'',k'')$ is its child then $|V(G'')|<|V(G')|$ by
description of the algorithm. It follows that each path from the
root to a leaf in the search tree has length $O(n)$. Considering the
statement of Lemma \ref{numleaves}, we get that the search tree has
$O(4^kn)$ nodes. The runtime of $FindCut(G,X,Y,k)$ can be
represented as a number of nodes of the search tree multiplied by
the runtime spent by the algorithm \emph{per} node. The heaviest
operations performed by the algorithm at the given node $(G,X,Y,k)$
are checking whether $Sep(G,x_l,Y)>k$ and, if not, checking whether
$Sep(G^C(u),x_l,Y)=Sep(G,x_l,Y)$ for a particular leaving neighbor
$u$ of $x_l$. Clearly these operations can be performed in a time
polynomial in $n$, where the degree of the polynomial is a constant
independent on $k$ (by applying a network flow algorithm). Thus the
runtime of $FindCut(G,X,Y,k)$ is $O(4^k*poly(n))$. Since the input
graph $G_{IN}$ may not satisfy our assumptions regarding the
minimality of the vertices of $X$ and the maximality of the vertices
of $Y$, the entire algorithm for the \textsc{ord-mc-dag} problem
includes also the transformation shown in the beginning of the
section. However the transformation can be performed in a polynomial
time and hence is taken into consideration by the expression
$O(4^k*poly(n))$. Thus we have proved the following theorem.

\begin{theorem} \label{part2}
There is an FPT-algorithm solving the parameterized
\textsc{ord-mc-dag} problem in time $O(4^k*poly(n))$
\end{theorem}

Theorem \ref{main} immediately follows from the combination of
Theorems \ref{part1} and \ref{part2}.

\section*{Acknowledgements}
We would like to thank Jianer Chen and Songjian Lu for providing a
copy of their WADS 2007 paper \cite{ChenLiu}.
\bibliographystyle{plain}
\bibliography{constr-doc}
\end{document}